\documentclass[seceq]{ptptex}

\usepackage{graphicx}
\usepackage{wrapft}

\title{
Boundary states in the open string channel and CFT near a corner %
}

\author{
Yosuke \textsc{Imamura}$^{1,}$\footnote{E-mail: imamura@hep-th.phys.s.u-tokyo.ac.jp},  
Hiroshi  \textsc{Isono}$^{2,}$\footnote{E-mail: isono@hep-th.phys.s.u-tokyo.ac.jp}
and Yutaka \textsc{Matsuo}$^{3,}$\footnote{E-mail: matsuo@phys.s.u-tokyo.ac.jp}
}

\inst{
Department of Physics, Faculty of Science, University of Tokyo \\
Hongo 7-3-1, Bunkyo-ku, Tokyo 113-0033, Japan
}

\abst{
We generalize the idea of boundary states to the open string
channel.  They describe  emission and absorption 
of open strings in the presence of intersecting D-branes.
We construct the explicit oscillator representation 
for the free boson and  fermionic ghost.
The inner product of such states describes a disk amplitude
of rectangular shape and possesses modular covariance
with a nontrivial conformal weight.
We compare the result obtained here with those obtained using
two different methods, one employing 
the path integral formalism and one employing the conformal anomaly.
We find that all these methods give consistent results.
In our method, we must be careful
in our treatment of the singularity of the CFT near the corners.
Specifically, we derive the correction to the conformal weight
of the primary field inserted at the corner, and it gives
the modular weight of the rectangle amplitude.
We also carry out explicit computations of the 
correlation functions.
}

\newcommand{\wt}{\widetilde}
\def\SL{\mathop{\rm SL}}
\newcommand{\wh}{\widehat}
\def\tr{\mathop{\rm tr}\nolimits}

\begin{document}

\maketitle


\section{Introduction}\label{intro.sec}
A boundary state \cite{r:BCFT,Polchinski:1987tu}
is one of the most fundamental
objects in the boundary conformal field theory.
It encodes the effect of the boundary conditions
in the operator formalism on a 2D world sheet.
In string theory, it gives an exact description 
of D-brane by specifying how
a closed string is emitted or absorbed
by the D-brane.

In the language of 2D conformal field theory,
the boundary state is characterized by the relation
\begin{equation}\label{bc1}
T_{\sigma\tau}|B^c\rangle=0,
\end{equation}
or in the Fourier modes of the Virasoro algebra,
\begin{equation}\label{bc2}
 (L_n-\tilde L_{-n})|B^c\rangle =0\,.
\end{equation}
The superscript $c$ is attached to indicate that 
this boundary state belongs to closed string Hilbert space.
There have been many works studying various properties of such states.
(For reviews, see, for example, Refs.~\citen{9912161,9912275,Behrend:1999bn,
Fuchs:1999zi,Frohlich:1999ss,Gaberdiel:2000jr,Billo:2000yb,
Schomerus:2002dc}).

As far as we know, the boundary state condition 
(\ref{bc1}) has been considered only in the closed string channel.
Such studies are natural, because the D-brane acts as a source of
emission/absorption of closed strings, and the boundary state
describes such a process.  For example, the inner product
of the boundary states $\langle B^c|q^{L_0+\tilde L_0-\frac{c}{12}}
|B^c\rangle$ describes closed string propagation between
D-branes.
The contribution of massless closed string fields to this amplitude
is identical to the potential derived using
supergravity, and with this information, we can determine
the tensions of D-branes. \cite{r:Polchinski}

As we see below, however,
it is also possible to consider a similar process
for open strings.
We consider the situation in which
two D-branes $\Lambda$ and $\Sigma$
intersect. \footnote{
Strictly speaking, in the analysis given in this paper,
it is not necessary that $\Lambda$ and $\Sigma$ 
``intersect'' in the usual geometrical sense. 
In fact, it is possible to define the open boundary state 
even if $\Lambda$ (resp. $\Sigma$)
is embedded in $\Sigma$ (resp. $\Lambda$) or
$\Lambda {\rm and }\Sigma$ are the same brane.
The only condition which we need is that $\Sigma\cap \Lambda$ is not the null set.
}
\begin{figure}[hbt]
	\begin{center}
	\scalebox{1.0}[1.0]{\includegraphics{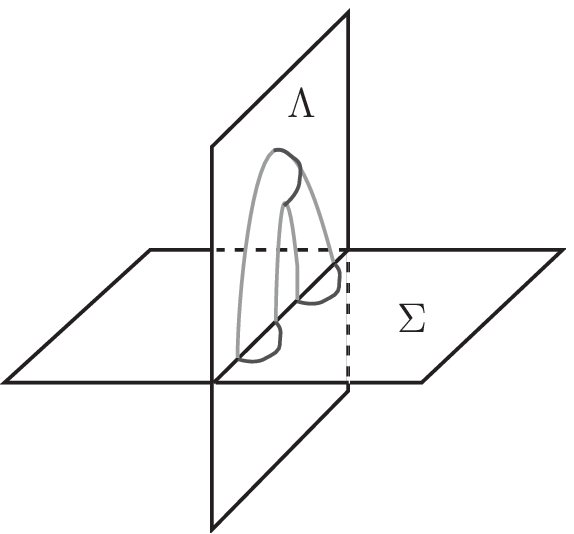}}
	\end{center}
	\caption{An open string on the D-brane $\Lambda$ is created/absorbed 
by another D-brane, $\Sigma$.}
	\label{fig:open_BS}
\end{figure}
Consider the physical process consisting of 
(1) an open string on a D-brane $\Lambda$ being emitted from a D-brane $\Sigma$,
(2) this open string propagating on the world volume of  $\Lambda$ and, 
(3) it being absorbed by $\Sigma$ (see Fig.\ref{fig:open_BS}).
As the closed string amplitude,
such a process would be described by an inner product of
the boundary state as
\begin{equation}\label{rect}
 {}_{\Lambda\Lambda}^{\Sigma}\langle B^o| q^{L_0^{
\Lambda\Lambda}+a^{\Lambda\Lambda}}|B^o\rangle^{\Sigma}_{\Lambda\Lambda}\,,
\end{equation}
where $|B^o\rangle^{\Sigma}_{\Lambda\Lambda}$ 
belongs to the open string Hilbert space
with both ends on the D-brane $\Lambda$
and $a^{\Lambda\Lambda}$ is the zero-point energy for 
the $\Lambda\Lambda$ sector.
We refer to such states as {\em open boundary states} (OBS)
\footnote{A partial result on OBS is announced in
Ref.~\citen{Isono:2005ie}.}. The first purpose of this paper is to define
such states and to give their explicit representation
for free fields.

We next study some basic properties of
such states. Firstly,
the inner product given by (\ref{rect}) gives a 
string amplitude with rectangular worldsheet.  In general, it is possible to use
different boundary conditions at the four sides.
While this worldsheet has the topology of a disk, 
it possesses a modular property that can be written schematically as
\begin{equation}
\label{modular}
 {}_{\Sigma_b,\Sigma_t}^{\Lambda_r}\langle B^o| 
\tilde q^{L_0^{\Sigma_t\Sigma_b}+a^{\Sigma_t\Sigma_b}}
|B^o\rangle_{\Sigma_t,\Sigma_b}^{\Lambda_l}
= t^{w(\Sigma_t|\Lambda_l,\Lambda_r|\Sigma_b)}\ 
 {}_{\Lambda_r,\Lambda_l}^{\Sigma_t}\langle B^o|  
q^{L_0^{\Lambda_l\Lambda_r}+a^{\Lambda_l\Lambda_r}}
|B^o\rangle_{\Lambda_l,\Lambda_r}^{\Sigma_b}\,,
\end{equation}
where $q=e^{-\pi t} and \tilde{q}=e^{-\pi /t}\,,$
for a real $t>0$. The symbol
$|B^o\rangle_{\Lambda_l,\Lambda_r}^{\Sigma_b}$
represents OBS with two independent boundary conditions,
$\Lambda_{r} {\rm and }\Lambda_{s},$ at two sides.  
The quantity ${w(\Sigma_t|\Lambda_l,\Lambda_r|\Sigma_b)}$
is the modular weight associated with the rectangle. 
This modular property implies that  the rectangle amplitude is described 
by modular forms, in contrast to the usual disk amplitude.

It is rather intriguing that
the disk amplitude may have such a modular covariance.
We present two arguments from which this can be understood.
The first is the most straightforward one, namely that regarding 
the evaluation of the path integral on the rectangle.  Although we
need to introduce a regularization, as usual, the amplitude is
manifestly covariant, by definition, and is identical with 
the inner product of OBS.

Another argument, which is more illuminating,
is that the rectangle amplitude is, strictly 
speaking,  the same as the disk amplitude {\em with} 
four marked points on the boundaries.  
These points are mapped to the four corners
of the rectangle. There is a relation between the cross ratio 
of the marked points with the modulus parameter $t$ defined above.
The conformal properties of CFT near the corner play
an essential role. Specifically, we show that the modular weight 
${w(\Sigma_t|\Lambda_l,\Lambda_r|\Sigma_b)}$
is the sum of the contributions from the corners.

We observe that the general behavior of CFT near the corner
with a deficit angle results from the conformal anomaly of the
stress-energy tensor.  After we map from the
disk amplitude, each corner has a primary
insertion with a weight proportional
to the central charge.  When the boundary conditions
of the neighboring sides are different, we also need
to add a twist field. The modular weight 
${w(\Sigma_t|\Lambda_l,\Lambda_r|\Sigma_b)}$
is proportional to the sum of the conformal weights 
of the inserted fields at each corner.
We also must be careful in choosing the normal ordering of operators
inserted at the corner.  We show that near the corner, 
each operator effectively becomes two operators, as its mirror image also acts.
We show this through explicit computation of the correlation function.

The organization of this paper is as follows.
In \S\ref{const.sec}, we give the definition of the 
boundary state in the open string channel.  
We then compute various rectangle amplitudes for the free
boson case and derive their modular properties.
In \S\ref{weight.sec}, in order to compare these amplitudes 
with those obtained using path integral computation, 
we study the general behavior of CFT at the corner with a deficit angle.
In \S\ref{path.sec},
we carry out the path integral evaluation of the rectangle
amplitude. We present two equivalent computations,
that employing the direct computation of the Laplacian on the rectangle
and the derivation from the conformal anomaly.
We find that both results coincide with that of the operator calculation
through the open boundary states.
In \S\ref{example.sec}, we evaluate four point
functions on the rectangle as an application
of our formulation.  
These analyses are sufficient to establish
various properties of the open boundary states.
 
After we submitted the first version of this paper to the arXiv, 
we were informed that the content of 
Ref.~\citen{Ilderton:2004xq} partially overlaps
that of this paper (in \S 2 and \S\S 4.2).
We improved the treatment of the stress energy tensor given in \S 2,
following this reference.

\section{Construction of the open boundary state}\label{const.sec}
\subsection*{\bf General formula}
The definition of the open boundary state can be derived from
the closed string sector (\ref{bc2}) using the doubling technique.
Let us consider the energy momentum tensor as the first example.
We consider an open worldsheet of rectangular shape
($0\leq \sigma\leq 2\pi$, $\tau\geq 0$). We identify the
anti-holomorphic part of the energy-momentum tensor as
\begin{equation}
 \bar T(\bar w)=T(2\pi-\bar w)\,, \quad
w=\sigma+i\tau\,,
\end{equation}
which gives the condition at $w=\pi$.
The combined tensor field $T(w,\bar w)=T(w)+\bar T(\bar w)$
then satisfies the boundary condition (\ref{bc1}) at the boundaries
$\sigma=0$ and $\sigma=\pi$.  
The boundary condition at $\tau=0$ is 
\begin{equation}\label{bc4}
(T(\sigma)-T(-\sigma))|B^o\rangle=0\,
\end{equation}
for $0<\sigma<\pi$. We see 
below that care is needed at $\sigma=0$ and $\sigma=\pi$
and indeed an anomaly exists there.
This condition is reminiscent of the definition of
the identity state in open string field theory \cite{Gross:1986ia},
\begin{equation}
(T(\sigma)-T(\pi-\sigma))|I\rangle =0\,.
\end{equation}
In a sense, the relation between the OBS and the identity state is 
similar 
to that between the boundary state in the closed string channel and
the crosscap state \cite{r:BCFT}:  
$|B^c\rangle \leftrightarrow |B^o\rangle$,
$|C^c\rangle \leftrightarrow |I\rangle$.

In order to define the open boundary state, we have to specify
three boundary conditions, two conditions at $\sigma=0,\pi$
and one condition at $\tau=0$.  The first two,
in general, are implemented by the doubling technique
(for generic conformal fields $\phi$ and $\bar\phi$), represented by
\begin{equation}\label{r}
 \bar\phi_\alpha(\bar w)=(\Lambda_r)_{\alpha\beta}\phi_\beta(2\pi-\bar w)\,,
\end{equation}
and the periodicity of the chiral field $\phi_\alpha$,
\begin{equation}\label{G}
 \phi_\alpha(w+2\pi)=G_{\alpha\beta}\phi_\beta(w)\,.
\end{equation}
Here, $\Lambda_r$ or $G$ are complex matrices.
The boundary condition at $\sigma=0$ is then specified by
\begin{equation}\label{l}
 \bar\phi_\alpha(\bar w)=
(\Lambda_{r} G)_{\alpha\beta}\phi_\beta(-\bar w)
\equiv (\Lambda_{l})_{\alpha\beta}\phi_\beta(-\bar w)\,.
\end{equation}
This determines the monodromy matrix $G$
in terms of $\Lambda_{l,r}$
\begin{equation}\label{monodr}
 G=(\Lambda_r)^{-1}\Lambda_l\,.
\end{equation}
The boundary condition at $\tau=0$ can be defined similarly as
\begin{equation}\label{b}
 \bar \phi_\alpha(\bar w) =
(\Sigma_b)_{\alpha\beta} \phi_\beta(\bar w).
\end{equation}
We need some consistency conditions in order to 
impose these conditions.  For example, the consistency condition 
obtained by considering the vicinity of $w=0$ is
\begin{eqnarray}
 \bar \phi(-\bar w)&=&\Sigma_b\cdot \phi(-\bar w)
=\Sigma_b \cdot(\Lambda_l)^{-1}\cdot\bar\phi(\bar w)\nonumber\\
&=& \Lambda_l\cdot\phi(\bar w)= 
\Lambda_l \cdot (\Sigma_b)^{-1}\cdot \bar\phi(\bar w)\,,
\end{eqnarray}
which implies
\begin{equation}\label{bl1}
 \Sigma_b (\Lambda_l)^{-1}=\Lambda_l (\Sigma_b)^{-1}
\quad\mbox{or}\quad  ( \Sigma_b (\Lambda_l)^{-1})^2=1\,.
\end{equation}
Similarly, the consistency condition obtained by considering 
the vicinity of $w=\pi$ is
\begin{equation}\label{br1}
  \Sigma_b (\Lambda_r)^{-1}=\Lambda_r (\Sigma_b)^{-1}
\quad\mbox{or}\quad  ( \Sigma_b (\Lambda_r)^{-1})^2=1\,.
\end{equation}
The definition of the boundary state can be written
in terms of the chiral field alone. There are two
equivalent ways to write the constraint on
$|B^o\rangle_{\Lambda_{l}\Lambda_{r}}^{\Sigma_{b}}$,
\begin{eqnarray}
&&\left(\phi_\alpha(-\sigma)-
(\Lambda^{-1}_l\Sigma_b)_{\alpha\beta}\phi_\beta(\sigma)
\right)|B^o\rangle_{\Lambda_{l}\Lambda_{r}}^{\Sigma_{b}}=0\,,
\nonumber\\
&&\left(\phi_\alpha(2\pi-\sigma)-
(\Lambda^{-1}_r\Sigma_b)_{\alpha\beta}\phi_\beta(\sigma)\right)
|B^o\rangle_{\Lambda_{l}\Lambda_{r}}^{\Sigma_{b}}=0\,.
\end{eqnarray}
Their equivalence can be shown trivially using (\ref{monodr}).

\subsection*{\bf Free bosons without a twist}
Let us use a free boson (without a twist) 
in order to obtain the simplest explicit formula.
The possible boundary conditions are those of the Neumann
and Dirichlet types. In terms of $J(w)=\partial X(w)$, they
can be written as
\begin{equation}
 J(-\bar w)+\epsilon \bar J(\bar w)= 0\,,\quad
\mbox{at Re}(w)=0\,.
\end{equation}
Here we have $\epsilon=1$ for Neumann and $\epsilon=-1$ for 
Dirichlet boundary condition.  Employing the same notation as in
our general discussion, the reflection matrices
are identified as $\Lambda_{r,l}=-\epsilon$.
Because we have $G=\epsilon_l\epsilon_r$, the chiral field is
periodic when $\epsilon_l=\epsilon_r$ and anti-periodic when
$\epsilon_l=-\epsilon_r$.
The well-known free boson mode expansions are,
\begin{eqnarray}
&&X^{(NN)}(w,\bar w)=\hat x-\alpha' \hat p(w-\bar w)
+i\left(\frac{\alpha'}{2}\right)^{1 \over 2} \sum_{m\neq 0}\frac{1}{m}
\alpha_m(e^{imw}+e^{-imw})\,,\\
&&X^{(DD)}(w,\bar w) = 
x+\frac{y-x}{2\pi}(w+\bar w)+
i\left(\frac{\alpha'}{2}\right)^{1 \over 2} \sum_{m\neq 0}\frac{1}{m}
\alpha_m(e^{imw}-e^{-imw})\,,\\
&& X^{(DN)}(w,\bar w) =x +i\left(\frac{\alpha'}{2}\right)^2
\sum_{r\in Z+1/2}\frac{1}{r}\alpha_r(e^{irw}-e^{-ir\bar w})\,,\\
&& X^{(ND)}(w,\bar w) =x +i\left(\frac{\alpha'}{2}\right)^2
\sum_{r\in Z+1/2}\frac{1}{r}\alpha_r(e^{irw}+e^{-ir\bar w})\,.
\end{eqnarray}
The commutation relations for the mode variables are
\begin{eqnarray}
 [\alpha_n,\alpha_m]=n\delta_{n+m,0}\,,\quad
 [\hat x,\hat p]=i\,.
\end{eqnarray}
We note that the zero mode sector is nontrivial only for NN.

The condition for the boundary state can be written 
in a form similar to that for the closed string case:
\begin{equation}\label{bc3}
 \partial_\tau X(\sigma,\tau)|_{\tau=0}|B^o\rangle^{N}_{\Lambda_l\Lambda_r}
=0\,,\qquad
 \partial_\sigma X(\sigma,\tau)|_{\tau=0}
|B^o\rangle^{D}_{\Lambda_l\Lambda_r}=0\,.
\end{equation}
The only difference between the open and closed string cases
is that the mode expansion depends on the boundary
conditions for open strings and we have only 
one set of oscillators instead of two (the left and right mover).

The open boundary states are obtained by solving (\ref{bc3}).
The result can be written compactly as
\begin{eqnarray}
|B^o\rangle^{\epsilon_b}_{\epsilon_l\epsilon_r}
&\propto& \exp\left(
-\epsilon_l\epsilon_b\sum_{n>0} \frac{1}{2n}
\alpha_{-n}^2
\right)|\mbox{zero mode}\rangle_{\epsilon_l,\epsilon_r}^{\epsilon_b}\,.
\end{eqnarray}
Here, $\epsilon_{l,r,b}$ represent the boundary conditions
at the left, right and bottom, respectively, and are $+1$ for Neumann boundary
conditions and $-1$ for Dirichlet boundary conditions.
The sum is over all positive integers when $\epsilon_l=\epsilon_r$
and all half odd positive integers when $\epsilon_l\neq \epsilon_r$.
The vacuum state $|\mbox{zero mode}\rangle_{\epsilon_l,
\epsilon_r}^{\epsilon_b}$ is the simple Fock vacuum when 
$(\epsilon_l,\epsilon_r)\neq (+1,+1)$. For 
$(\epsilon_l,\epsilon_r)= (+1,+1)$, we need to prepare
the zero mode wave function.  An appropriate choice is
$|\hat p=0\rangle$ for $\epsilon_b=1$ and
$|\hat x=x_0\rangle$ with $x_0\in \boldsymbol{R}$ for $\epsilon_b=-1$.

\subsection*{\bf Anomalies at the corners}
Let us study the condition (\ref{bc4}) in our specific constructions.
It would be replaced by the condition
\begin{equation}\label{naive}
 (L_n-L_{-n})|B\rangle^{\epsilon_b}_{\epsilon_l\epsilon_r}\stackrel{?}{=} 0\,.
\end{equation}
It turns out, however, that we have an extra term if we apply the Virasoro
generators ($n\neq 0$) in terms of oscillators,
\begin{equation}
 L_n=\frac{1}{2}\sum_{m}\alpha_{n-m}\alpha_{m}
\end{equation}
where $m$ runs over all integers for NN and DD boundary conditions
and all half odd integers for DN and ND boundary conditions.
With this extra term, we obtain, instead of (\ref{naive}),
\begin{equation}\label{anomaly_corner_L}
 (L_n-L_{-n}+(d_n)_{\epsilon_l\epsilon_r}^{\epsilon_b})
|B\rangle^{\epsilon_b}_{\epsilon_l\epsilon_r}=0\,,
\end{equation}
where
\begin{equation}
(d_n)_{\epsilon_l\epsilon_r}^{\epsilon_b}=-2n
(\lambda_l+(-1)^n \lambda_r)\,,
\quad 
\mbox{with}\quad\lambda_l=-\frac{\epsilon_b\epsilon_l}{16}
\,,\,
\lambda_r=-\frac{\epsilon_b\epsilon_r}{16}\,.
\end{equation}
We show below that the quantities $\lambda_{l,r}$ are the conformal
weights of the operators inserted at the corners.
The meaning of this anomaly term becomes clearer if
we write it as a condition for $T(\sigma)$ :
\begin{equation}\label{anomaly_corner_T}
\left(T(\sigma)-T(-\sigma)+4\pi i
\left(\lambda_l\delta'(\sigma)
+\lambda_r\delta'(\sigma-\pi)\right)\right) 
|B\rangle^{\epsilon_b}_{\epsilon_l\epsilon_r}=0\,.
\end{equation}

\subsection*{\bf Free bosons twisted by flux}
One of the most straightforward generalizations of free boson OBS
is the inclusion of the fluxes. In the case of a closed string, 
the boundary condition becomes (for example, see Ref.~\citen{9912275})
\begin{equation}
  \left(\partial_{\bar z}X^\mu
-\mathcal{O}^\mu_\nu \partial_z X^\nu\right)|B\rangle^{\mathcal{O}}=0\,,\quad
{\mbox{where }}\mathcal{O}=(1+F)(1-F)^{-1}\,.
\end{equation}
Here, $\mathcal{O}$ satisfies the orthogonality condition
$\mathcal{O}\cdot\mathcal{O}^t=1$ and is an element of ${\rm O}(d)$
(where $d$ is the number of dimensions in which the flux $F_{\mu\nu}$
is introduced).

In order to define OBS, we have to specify three D-branes,
which are located on the left, right and bottom.  We consider
the situation in which these D-branes have the fluxes $F_l$, $F_r$
and $F_b$, and we denote the orthogonal matrices associated
with these fluxes as $\mathcal{O}_l$, $\mathcal{O}_r$ and $\mathcal{O}_b$,
respectively.
The identifications of the holomorphic and anti-holomorphic
oscillators at the three sides become
\begin{eqnarray}
&&\left.\left( \partial_{\bar z}X^\mu
+(\mathcal{O}_l)^\mu_\nu \partial_z X^\nu\right)\right|_{\sigma=0}=0
\,,\qquad
\left.\left( \partial_{\bar z}X^\mu
+(\mathcal{O}_r)^\mu_\nu \partial_z X^\nu\right)
\right|_{\sigma=\pi}=0\,,\\
&&  \left(\partial_{\bar z}X^\mu
-(\mathcal{O}_b)^\mu_\nu \partial_z X^\nu\right)
|B^o\rangle^{\mathcal{O}_b}_{\mathcal{O}_l\mathcal{O}_r}=0\,.
\end{eqnarray}
The two conditions on the first line are used to determine
the identification of the anti-holomorphic part with the holomorphic part
and fix the periodicity of the latter
under the shift $\sigma\rightarrow \sigma+2\pi$.
The condition on the second line represents
the explicit definition of OBS in terms of holomorphic
oscillators. 

These constraints have the same form as those used in the general discussion
if we make the replacements 
$\Sigma_b\rightarrow \mathcal{O}_b$, 
$\Lambda_r\rightarrow -\mathcal{O}_r$ and 
$\Lambda_l\rightarrow \mathcal{O}_l$,
and hence we can use the consistency conditions derived there.
In the present case they can be written as
\begin{equation}\label{blr}
  \mathcal{O}_b^T \mathcal{O}_l=\mathcal{O}_l^T \mathcal{O}_b\,,\quad
  \mathcal{O}_b^T \mathcal{O}_r=\mathcal{O}_r^T \mathcal{O}_b\,.
\end{equation}

We do not attempt to write explicitly all the matrices
that solve these constraints.  However, it is rather easy to see
that there are many of solutions.  A trivial solution is
$\mathcal{O}_{b,l,r}=\epsilon_{b,l,r}\mathcal{O}$, where
$\mathcal{O}\in {\rm O}(d)$ and $\epsilon_{b,l,r}=\pm 1$.
This gives a D-brane with the same flux or its $T$-dual.

For a more nontrivial solution, let us note that if we define
$A=\mathcal{O}^T_r\mathcal{O}_b$, then we have $A\in {\rm O}(d)$, and
the constraint (\ref{blr}) implies $A^2=1$.
This is not so difficult to solve for small values of $d$.
The simplest nontrivial case is $d=2$, for which there is a 
one-parameter family for $A$,
\begin{equation}
A=\mathcal{O}^T_r\mathcal{O}_b=
\left(
\begin{array}{c c}
  \cos\theta& \sin\theta\\ \sin\theta & -\cos\theta
\end{array}
\right)
=
\left(
\begin{array}{c c}
  \cos\theta_1& -\sin\theta_1\\ \sin\theta_1 & \cos\theta_1
\end{array}
\right)
\left(
\begin{array}{c c}
  \cos\theta_1& \sin\theta_2\\ \sin\theta_2 & -\cos\theta_2
\end{array}
\right)\,,
\end{equation}
with $\theta=\theta_1+\theta_2$.
This implies that there is a large class of solutions of the form
\begin{eqnarray}
&&  \mathcal{O}_r=\left(
\begin{array}{c c}
  \cos\theta_1& -\sin\theta_1\\ \sin\theta_1 & \cos\theta_1
\end{array}
\right)\,,\,
  \mathcal{O}_l=\left(
\begin{array}{c c}
  \cos\theta_2& -\sin\theta_2\\ \sin\theta_2 & \cos\theta_2
\end{array}
\right)
\,,\,\nonumber\\
&&  \mathcal{O}_b=\left(
\begin{array}{c c}
  \cos\theta_3& \sin\theta_3\\ \sin\theta_3 & -\cos\theta_3
\end{array}
\right)\,,
\end{eqnarray}
for arbitrary $\theta_{1,2,3}\in \boldsymbol{R}$.

The interpretation of this solution is straightforward.
Suppose we choose $\theta_3=0$.  The $\mathcal{O}_b$ in this case 
describes a D1-brane in 2 dimensions.  The angle $\theta_3$ will rotate
this D1-brane in an arbitrary direction.
On the other hand, the D-branes on the left and right are
D2-branes with arbitrary flux.  We note that there are also
solutions for which the left and right D-branes are D1 branes at arbitrary
angles and the bottom one is a D2-brane with arbitrary flux.

Once the constraints (\ref{blr}) are solved, the mode expansion
of $X$ is determined so as to satisfy the periodicity condition
\begin{equation}
 \partial X(\sigma+2\pi)=\mathcal{O}_r^T\mathcal{O}_l \partial X(\sigma)\,,
\end{equation}
and the boundary state is determined by the condition
\begin{equation}
 (\partial X(-\sigma)+\mathcal{O}^T_l \mathcal{O}_b \partial X(\sigma))
|B^o\rangle^{\mathcal{O}_b}_{\mathcal{O}_l \mathcal{O}_r}=0\,.
\end{equation}
While these expressions are rather formal, it is not difficult to
confirm that these definitions are consistent.

\subsection*{\bf Ghosts}
For the investigation of a bosonic string, 
we need the open boundary state for the reparametrization ghosts.
Unlike the Majorana fermions, which turn out to be more nontrivial,
the open boundary state for the ghost fields can be obtained in a straightforward manner. 
Here, the boundary conditions become,
\begin{equation}
 (b_n-b_{-n})|B^o\rangle^{\rm (gh)}= (c_n+c_{-n})|B^o\rangle^{\rm(gh)}=0
\,.
\end{equation}
These conditions can be solved immediately as
\begin{equation}
 |B^o\rangle^{\rm (gh)}=
\exp\left(\sum_{n>0}c_{-n}b_{-n}\right)
c_0 c_{1}|0\rangle^{\rm (gh)}\,,
\label{Bghdef}
\end{equation}
where $|0\rangle^{\rm (gh)}$ is the $\SL(2,\boldsymbol{R})$-invariant
ghost vacuum.

As mentioned above, the inner product between the
open boundary states describes a rectangular worldsheet . 
At the corners, we have a deficit angle of $\pi/2$, which leads to some
nontrivial features of the OBS.  We give a systematic
description for generic CFT in the next section.
Here we give a short description which is specific to the
ghost field. In this case, the effect of the deficit angle
appears as the (implicit) ghost insertion
at both ends of an open string.
This can be seen in the following way.
Let us focus on one of the two endpoints, say $\sigma=0$:
\begin{equation}
c(u)|B^o\rangle^{\rm (gh)}\sim{\cal O}(u),\quad
b(u)|B^o\rangle^{\rm (gh)}\sim{\cal O}(1).
\end{equation}
The deficit angle at the corner is mapped to a smooth boundary
by the conformal map $z=u^2$.  In terms of this new coordinate $z$,
we obtain the following  behavior of the $b$ and $c$ fields
in the upper half plane:
\begin{equation}
c(z)=2u c(u)\propto z,\quad
b(z)=(2u)^{-2}b(u)\propto\frac{1}{z}.
\label{bcatcorner}
\end{equation}
These relations imply that
in the state (\ref{Bghdef}), the ghost operator $c$ has
been inserted at $\sigma=0$ in the smooth coordinate $z$.
We interpret this as meaning that the corners are ``marked points'' on
the worldsheet where ghost fields are inserted.
They fix the symmetry associated with the conformal Killing vectors
and  reproduce the FP determinant.
We will return to this point in \S\ref{path.sec}.

\subsection*{\bf BRST property of OBS}
In string theory, a physical state condition is written in terms of a BRST operator. 
For the open string boundary state, it gives a nontrivial constraint, 
due to the anomaly at the corner (\ref{anomaly_corner_L}). 
In bosonic string theory, the BRST operator in the general background 
can be written as 
\begin{eqnarray}
Q_B=\sum_{n\in\boldsymbol{Z}}c_n L_{-n}^{\rm mat}
+\sum_{m,n\in\boldsymbol{Z}} \frac{m-n}{2}
:c_m c_n b_{-m-n}:-c_0\,,
\end{eqnarray}  
and the open boundary state for free bosons takes the form
\begin{eqnarray}
|B^o\rangle\equiv\prod_{\rm bosons} |B^o\rangle^{\epsilon_b}_{\epsilon_l\epsilon_r}
\otimes|B^o\rangle^{\rm (gh)}\,.
\end{eqnarray}
When we apply the BRST operator to this boundary state, the result is
\begin{eqnarray}
Q_B|B^o\rangle &=&
-\sum_{n=1}^{\infty}c_{-n} \sum_{\rm bosons}(d_n)
^{\epsilon_b}_{\epsilon_l\epsilon_r}|B^o\rangle
-\sum_{n=1}^{\infty} 3nc_{-2n}|B^o\rangle
\nonumber\\
&=&
-\sum_{n=1}^{\infty}c_{-n}n\left[
\left(\frac{1}{8}\sum_{\rm bosons} \epsilon_b\epsilon_l+\frac{3}{4}\right)
+(-1)^n
\left(\frac{1}{8}\sum_{\rm bosons} \epsilon_b\epsilon_r+\frac{3}{4}\right)
\right]\,,
\end{eqnarray}
where we have used (\ref{anomaly_corner_L}).
Hence, the BRST invariance $Q_B|B^o\rangle=0$ is equivalent to 
\begin{eqnarray}
\sum_{\rm bosons} \epsilon_b\epsilon_l=-6,~~
\sum_{\rm bosons} \epsilon_b\epsilon_r=-6\,.
\label{eeism6}
\end{eqnarray}

For a more general background, 
we expect that the Virasoro condition for the OBS
remains in same form (\ref{anomaly_corner_L}), since it
reflects the general behavior of the stress tensor at the corner
as we see in the next section. In terms of this language,
the above conditions are replaced by $\lambda_l=\lambda_r=3/8$, 
where $\lambda_{l}$ and $\lambda_{r}$
are the conformal weights of the operators inserted at the corners.
Usually, the operator that is inserted at the boundary
should have a conformal weight 1 in order to maintain conformal
invariance.  We argue in the next section that this deviation comes
from the anomaly associated with the deficit angle at the corner.

\subsection*{\bf Modular covariance of rectangular diagram}
We can check the consistency of these states by calculating their inner product
and compare it with the usual path integral formula.
This gives a disk amplitude in the form of a rectangle,
whose sides are specified by 
various boundary conditions. Although this is a disk diagram,
we conjecture that the modular invariance (\ref{modular}) holds.

For the free case, we can readily confirm (\ref{modular}).\footnote{
A useful formula is
$$
\langle 0|e^{q a^2/2}e^{\pm (a^\dagger)^2/2}|0\rangle=
(1-q)^{\mp 1/2}\,
$$
for a simple oscillator, with $[a,a^\dagger]=1$.
}
A straightforward computation gives
\begin{equation}\label{inner}
I_{(\epsilon_t|\epsilon_l,\epsilon_r|\epsilon_b)}(t)
\equiv
{}^{\epsilon_t}_{\epsilon_r\epsilon_l}\langle B|
q^{L^{\epsilon_l\epsilon_r}_0+a^{\epsilon_l\epsilon_r}}
|B\rangle^{\epsilon_b}_{\epsilon_l\epsilon_r}
=\frac{q^{a^{\epsilon_l\epsilon_r}}}{\prod_{n>0}
\sqrt{1-\epsilon_t\epsilon_b q^{2n}}}
I^{(\rm zero)}_{(\epsilon_t|\epsilon_l,\epsilon_r|\epsilon_b)}(t)\,,
\end{equation}
where $q=e^{-\pi t}$, and 
the zero-point energy $a^{\epsilon_l\epsilon_r}$ is given by
\begin{equation}
a^{\epsilon_l\epsilon_r}
=-\frac{1}{24}\quad\mbox{for $\epsilon_l\epsilon_r=+1$},\quad
a^{\epsilon_l\epsilon_r}
=\frac{1}{48}\quad\mbox{for $\epsilon_l\epsilon_r=-1$}.
\end{equation}
As in the previous case, $n$ in (\ref{inner}) runs over positive integers
for $\epsilon_l\epsilon_r=1$ and positive half odd integers
for $\epsilon_l\epsilon_r=-1$.
We denote the contribution of the zero mode as
$I^{(\rm zero)}_{(\epsilon_t|\epsilon_l,\epsilon_r|\epsilon_b)}$.
It is nontrivial when 
$\epsilon_l=\epsilon_r=+1$ (NN) and trivial
(i.e. $I^{(\rm zero)}_{(\epsilon_t|\epsilon_l,\epsilon_r|\epsilon_b)}=1$) in all other cases, 
because unless $\epsilon_l=\epsilon_r=+1$,
($\epsilon_l,\epsilon_r$)-strings have no zero modes. 
There are four cases corresponding to the former condition, $\epsilon_l=\epsilon_r=+1$,
but only one of them gives a nontrivial contribution,
\begin{equation}
I^{(\rm zero)}_{(D|N,N|D)}=\frac{1}{\sqrt{\alpha 't}},\quad
I^{(\rm zero)}_{(N|N,N|N)}=
I^{(\rm zero)}_{(N|N,N|D)}=
I^{(\rm zero)}_{(D|N,N|N)}=1.
\end{equation}
Here we have used the following zero mode conventions:\footnote{
We note that  in deriving $I^{(\rm zero)}_{(D|N,N|D)}=\frac{1}{\sqrt{\alpha^{\prime}t}}$, 
we set the position eigenvalues of the zero mode
all equal, with the value $x_0$. 
If they are not all equal, we have to replace $I^{(\rm zero)}_{(D|N,N|D)}$ by
\begin{equation}
\langle x_0|q^{\alpha^{\prime}\wh p^2}|y_0\rangle
=\Biggl(-\frac{\pi}{\alpha^{\prime}\ln q}\Biggr)^{\frac{1}{2}}
\exp{\Biggl(\frac{(x_0-y_0)^2}{\alpha^{\prime}\ln q}\Biggr)}.
\end{equation}
}
\begin{eqnarray}
&{}&
\wh x=\frac{i}{2}(\wh a-\wh a^{\dagger}),~\wh p=\wh a+\wh a^{\dagger},~
[\wh a,\wh a^{\dagger}]=1 \nonumber\\
&{}&
|x_0\rangle=\Biggl(\frac{2}{\pi}\Biggr)^{\frac{1}{4}}
e^{\frac{1}{2}\wh a^{\dagger 2}-2ix_0\wh a^{\dagger}-x_0^2}|0\rangle,~
|p_0\rangle=(2\pi)^{\frac{1}{4}}
e^{-\frac{1}{2}\wh a^{\dagger 2}+p_0\wh a^{\dagger}-\frac{1}{4}p_0^2}|0\rangle \\
&{}&
\wh p|p_0\rangle=p_0|p_0\rangle,~\wh x|x_0\rangle=x_0|x_0\rangle \nonumber
\end{eqnarray}
Including these zero mode contributions,
we obtain the following expression for the rectangle
amplitude for a free boson:
\begin{eqnarray}
&&I_{(N|N,N|N)}
=I_{(D|D,D|D)}
=I_{(N|D,D|N)}
=\frac{1}{\eta^{1/2}(t)},
\label{zdddd}\\
&&I_{(D|N,N|D)}
=\frac{1}{(\alpha 't)^{1/2}\eta^{1/2}(t)},\quad
I_{(D|N,N|N)}
=I_{(D|D,D|N)}
=\frac{\eta^{1/2}(t)}{\eta^{1/2}(2t)},\\
&&I_{(N|N,D|N)}
=I_{(D|N,D|D)}
=\frac{\eta^{1/2}(t)}{\eta^{1/2}(t/2)},\quad
I_{(N|N,D|D)}
=\frac{\eta^{1/2}(2t)\eta^{1/2}(t/2)}{\eta(t)}\,.
\end{eqnarray}

As a consistency check, we study the modular property
of these partition functions. We can write this property as
\begin{equation}
I_{(\epsilon_1|\epsilon_2,\epsilon_4|\epsilon_3)}(1/t)
=t^w
I_{(\epsilon_4|\epsilon_1,\epsilon_3|\epsilon_2)}(t)\,,
\label{rectymodular}
\end{equation}
where $w$ is some weight which may depend on the boundary conditions.
By using the explicit forms given above,
we can confirm  this relation with the weight
\begin{equation}
w=-\frac{1}{16}(
\epsilon_1\epsilon_2+
\epsilon_2\epsilon_3+
\epsilon_3\epsilon_4+
\epsilon_4\epsilon_1)\,.
\label{w122}
\end{equation}
These factors have a structure  similar to that of the coefficients
of the anomaly term in (\ref{anomaly_corner_T}).  In the next section,
we see that they are indeed closely related.

We can make a similar calculation for the ghost sector.
The rectangle amplitude for the ghost field is
\begin{equation}
I_{\rm ghost}(t)={}^{\rm (gh)}\langle B^o| b_0 q^{L^{\rm (gh)}_0
+\frac{1}{12}}|B^0\rangle^{\rm (gh)}
=\eta(t),
\label{ghamp}
\end{equation}
and its modular property is
\begin{equation}
I_{\rm ghost}(1/t)
=t^{1/2}I_{\rm ghost}(t)\,.
\label{ghostmodular}
\end{equation}
This is the square root of the contribution
of the ghost boundary state in the closed string.

\subsection*{\bf A consistency condition for a bosonic string}
Let us consider a bosonic string theory with flat
intersecting D-branes. In this case, the rectangle amplitude
can be written as the product
$I^{\rm tot}=I_{\rm ghost}\prod_{\rm bosons}I$,
where
$\prod_{\rm bosons}$ represents the product over $26$ boson fields.
{}From (\ref{rectymodular}) and (\ref{ghostmodular}),
the modular property of $I^{\rm tot}$ is
\begin{equation}
I^{\rm tot}(1/t)=t^{w_{\rm tot}}I^{\rm tot}(t)\,,
\end{equation}
where the total weight $w_{\rm tot}$ is given by
\begin{equation}
w_{\rm tot}
=-\frac{1}{16}\sum_{\rm bosons}
\sum_{i=1}^4\epsilon_i\epsilon_{i+1}+\frac{1}{2}\,.
\label{wtot}
\end{equation}
The rectangle amplitude describes the propagation of an open string
between the two D-branes $\Sigma_b$ and $\Sigma_t$,
\begin{equation}
A=\int_0^\infty I^{\rm tot}_{(\Sigma_t|
\Lambda_l, \Lambda_r|\Sigma_b)}(t)dt\,.
\end{equation}
This amplitude can also be interpreted as the propagation from
$\Lambda_l$ and $\Lambda_r$ in the modular dual channel,
and $A$ should be invariant.
This implies that the total weight $w_{\rm tot}$ must be $2$.
Actually, we can show that this is the case
if the BRST constraint (\ref{eeism6}) is satisfied.

\section{CFT at the corner}\label{weight.sec}
In \S\ref{const.sec} we computed rectangle amplitudes with various boundary
conditions, and found that they are transformed
under modular transformations as modular forms.
The factor $t^w$ on the right-hand side of (\ref{rectymodular})
can be interpreted as a conformal factor arising in
the dilatation of the rectangular worldsheet from the size $1\times t$
to $1/t\times 1$. With this interpretation, each term
$-\epsilon_i\epsilon_{i+1}/16$
in the modular weight (\ref{w122}) seems to be the 
conformal weight associated with each corner of the rectangle.
In this section, we first derive the basic properties
of CFT near the corner, with special attention given to
the change of the conformal weight of the primary fields.\footnote{
For related studies of CFT at the corner, see, for example, 
Refs.~\citen{Cvetic:2003ch} and \citen{Abel:2003vv}.
}
We then prove that, for the rectangular diagram, 
the conformal weight at each corner indeed gives the modular
weight of the modular transformation of the rectangular diagram.

\subsection*{\bf CFT at the corner}
Let us focus on the vicinity of one corner and
use the first quadrant instead of a rectangle.
The map between the upper half plane with the coordinate $z'$
and the first quadrant with the coordinate $z$
is given by $z'=z^2$. In the following, we study the slightly  
more general case of $z'=z^n$.

Let ${\cal O}$ be a primary operator with conformal weight $\lambda$.
We insert this operator at the origin of the $z'$-plane. Then, at the origin,
the energy momentum tensor $T(z')$ has the double pole
singularity
\begin{equation}
T(z')=\frac{\lambda}{z'^2}+\cdots\,,
\label{tzp}
\end{equation}
where the dots represent a simple pole and the finite part.
For a general conformal map of the form $z'=f(z)$,
the energy momentum tensor in a theory with central charge $c$
is transformed according to
\begin{equation}
T(z)=(f')^2T(z')+c\left(\frac{f'''}{12f'}-\frac{(f'')^2}{8(f')^2}\right).
\end{equation}
Using this formula for the conformal map $z'=z^n$,
we can translate the behavior of $T(z')$ in (\ref{tzp})
to the following behavior of $T(z)$ near the corner:
\begin{equation}
T(z)=\left(n^2\lambda-c\frac{n^2-1}{24}\right)\frac{1}{z^2}+\cdots\,.
\end{equation}

In order to evaluate the conformal weight associated with
the corner, we have to divide the coefficient of $1/z^2$ by
$n$:\footnote{
This factor appears because
the dilatation operator near the edge is given by
the integral of $T(z)$ over the contour consisting of 
$1/n$ of the usual half-circle. More explicitly, 
this integral is taken over the path $z=r e^{i\theta}$
with $0\leq \theta\leq \pi/n$.
}
\begin{equation}
\lambda_{\rm corner}=n\lambda-c\frac{n^2-1}{24n}.
\label{lambdaeff}
\end{equation}
The first term on the right-hand side of this relation represents the contribution
of the inserted operator, and the second term is the intrinsic weight of the corner.

When boundary conditions on both sides of the corner
are the same (i.e.,DD or NN),
we need no operator insertion at the origin of $z'$, 
and we have $\lambda=0$.
In this case the equality $\lambda_{\rm corner}=-1/16$ holds for each free boson.
By contrast, if the two boundary conditions are different (ND or DN),
we need to insert the twist operator $\sigma$ for the boson field.
In this case, because $\sigma$ has conformal weight $1/16$,
Eq.(\ref{lambdaeff}) gives $\lambda_{\rm corner}=+1/16$.
These results can be combined to give
 $\lambda_{\rm corner}=-\epsilon_i\epsilon_{i+1}/16$,
which coincides with the modular weight of the rectangle
at one corner.
At the same time, it accounts for the anomaly at the corners
in (\ref{anomaly_corner_T}), if we note 
\begin{equation}
 \mbox{disc}\frac{1}{z^2}=-2\pi i\delta'(\sigma)\,.
\end{equation}

The change of the conformal weight of the primary fields
can be interpreted as the renormalization effect
induced by the conformal
map. We now rederive the first term in (\ref{lambdaeff}),
that is, the contribution of the inserted operator.
Let us consider the conformal map
$z'=z^n$ again.
Because the corner is a singular point,
we have to be careful when we insert an operator there.
To see what kind of singularity arises,
we first insert the operator at a generic point $z$,
and then take the limit $z\rightarrow0$.
The primary fields ${\cal O}(z)$ and ${\cal O}(z')$ are
related as
\begin{equation}
{\cal O}(z'=0)=\lim_{z\rightarrow 0}(nz^{n-1})^\lambda{\cal O}(z).
\label{ooprime}
\end{equation}
Because the $z'$-plane is smooth,
the insertion of ${\cal O}(z')$ does not produce
any singularity in the $z'\rightarrow 0$ limit.
This fact and the relation (\ref{ooprime}) together imply that
the operator ${\cal O}(z)$ in the $z$-plane
behaves as
$\sim(nz^{n-1})^\lambda$.
Therefore, in order to have a well-defined vertex insertion,
we should use the renormalized operator
\begin{equation}
{\cal O}^{\rm ren}(z)=\frac{1}{(nz^{n-1})^\lambda}{\cal O}(z).
\end{equation}
The renormalization factor
gives an extra contribution to the conformal weight.
The factor $(nz^{n-1})^{-\lambda}$
has conformal weight $(n-1)\lambda$, and
the total conformal weight of the renormalized operator is
$n\lambda$.
This is what we have in (\ref{lambdaeff}).

\subsection*{\bf Multiplication of tachyon vertices with OBS}
As a simple example, let us consider an
OBS with tachyon vertices inserted at the corners.
The (bare) tachyon vertex is
\begin{equation}
V_k(z)=:e^{-ikX(z)}:.
\label{vkdef}
\end{equation}
The conformal weight of this operator is $\lambda=\alpha^{\prime}k^2$.

We can explicitly see the singular behavior of this operator
near the corner $\sigma=0$ if we apply this operator to the OBS.
This is done below. We use NNN boundary conditions for concreteness.
First, Wick's theorem gives
\begin{eqnarray}
V(0,\tau)|B^o\rangle^N_{NN}
&=&
e^{-ik\wh x}
\prod_{n=1}^\infty\exp\left(-2\sqrt{2\alpha^{\prime}}k\frac{a_{-n}}{2n}e^{n\tau}\right)
\exp\left(2\sqrt{2\alpha^{\prime}} k\frac{a_n}{2n}e^{-n\tau}\right)
|B^o\rangle^N_{NN}
\nonumber\\
&=&
(1-e^{-2\tau})^{\alpha^{\prime} k^2}
e^{-ik\wh x}
\prod_{n=1}^\infty\exp\left(-2\sqrt{2\alpha^{\prime}} k\frac{a_{-n}}{n}\cosh(n\tau)\right)
|B^o\rangle^N_{NN}
\nonumber\\
&\sim&
(2\tau)^{\alpha^{\prime} k^2}
e^{-ik\wh x}
\prod_{n=1}^\infty\exp\left(-2\sqrt{2\alpha^{\prime}} k\frac{a_{-n}}{n}\right)
|B^o\rangle^N_{NN}\,.
\end{eqnarray}
Because of the factor $(2\tau)^{\alpha^{\prime} k^2}$,
this is singular at $\tau=0$.
An operator inserted near the other corner, $\sigma=\pi$,
produces a similar singularity.
In order to remove this singularity, we define the following renormalized
vertex operator:
\begin{equation}\label{rentach}
V^{\rm ren}_k(\tau,\sigma)=
(2\tau)^{-\alpha^{\prime} k^2}V_k(\tau,\sigma).
\end{equation}
This kind of renormalization is familiar in the relation
between the bulk operator and the boundary operator.
For example, a bulk operator $e^{ik\phi}$ has conformal weight
$(\alpha^{\prime}k^2/2,\alpha^{\prime}k^2/2)$.
Therefore, we could naively expect the conformal weight
$\alpha^{\prime}k^2/2+\alpha^{\prime}k^2/2=\alpha^{\prime}k^2$
for a boundary operator of the same form.
However, the actual value is $2\alpha^{\prime}k^2$.
This is because the limit in which the operator approaches the boundary
is singular, and therefore we need an extra renormalization factor, which
doubles the conformal weight.

\subsection*{\bf Localization of the weight at the corners}
\begin{figure}[t]
\begin{center}
\scalebox{1.0}[1.0]{\includegraphics{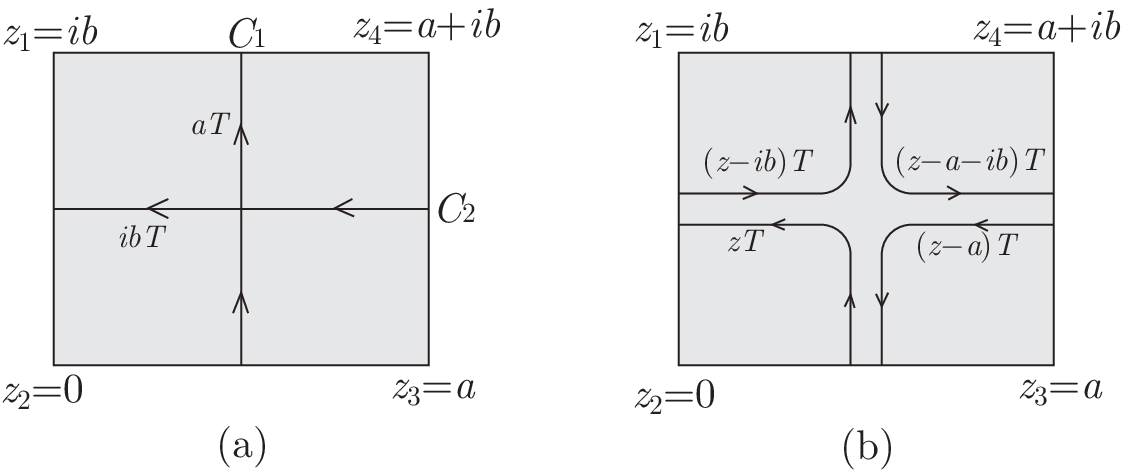}}
\end{center}
\caption{
(a)
The dilatation is generated by the energy momentum tensor
inserted along the two contours $C_1$ and $C_2$.
The operators integrated along each contour are shown beside that contour.
(b)
The two contours in (a) can be rearranged into four contours around each corner.
The integral along each contour gives the conformal weight of
the corresponding corner.
In both (a) and (b), we show only the holomorphic parts.
}
\label{weyl.eps}
\end{figure}

Finally, we can show the localization of the weight 
at the corners for the rectangle amplitude
in the following way.
Let us consider a rectangular worldsheet, and
let $a$ and $b$ denote the lengths
along the $\tau$ and $\sigma$ directions, respectively.
The deformation of the worldsheet is
generated by the energy momentum tensor.
For example, a change of the length $b$ to $b+\delta b$ is
generated by $\delta b L_0=\delta b\int_{C_2}\frac{d\sigma}{2\pi}T+$
[anti-holomorphic part],
where the contour $C_2$ is a segment crossing worldsheet from one boundary,
$\sigma=0$, to the opposite boundary, $\sigma=a$.
Similarly, the change of the length $a$ is generated by  the insertion of $T$
along the contour $C_1$, which goes from the boundary $\tau=0$ to $\tau=b$.
Combining these two, we can generate the dilatation of the
worldsheet, and the change of amplitude resulting from the dilatation is
given by
\begin{equation}
\frac{\delta I}{I}
=\left\langle\oint_{C_1}\frac{dz}{2\pi i}aT
+\oint_{C_2}\frac{dz}{2\pi i}ibT\right\rangle
+\mbox{c.c.}
\label{contourc1c2}
\end{equation}
where c.c. (complex conjugate) represents the contribution of
the anti-holomorphic part
[see Fig.\ref{weyl.eps} (a)].
We can deform the two contours into four contours going around each corner.
Doing so, we rewrite (\ref{contourc1c2}) as
\begin{equation}
\frac{\delta I}{I}=\sum_{i=1}^4\left\langle\oint_{z_i}\frac{dz}{2\pi i}(z-z_i)T\right\rangle+\mbox{c.c.},
\label{dzzsum4}
\end{equation}
where $z_i$ ($i=1,2,3,4$) are the coordinates of the four corners,
and $\int_{z_i}$ represents integration along the contour around $z_i$
[see Fig.\ref{weyl.eps} (b)].
In (\ref{dzzsum4}),
the integral for each $i$ extracts the conformal weight
of the $i$-th corner.
This argument implies that
the effect of the dilatation is obtained by summing
the conformal weights associated with the corners.

\section{Path integral approach to the rectangle amplitude}
\label{path.sec}
In this section, we present the path integral evaluation
of the rectangle amplitude and confirm that it is identical to the results 
obtained from OBS. 

\subsection{Direct evaluation of the path integral}
Formally, the amplitude is obtained from the Gaussian integral
\begin{equation}
I=\int{\cal D}X e^{-S[X]}=(\det\Delta)^{-1/2},
\end{equation}
where $S[X]$ is the action of the matter part,
\begin{equation}
S[X]=\frac{1}{4\pi\alpha'}\int d^2\sigma X\Delta X\,
\end{equation}
and $\Delta$ is (up to a sign) the Laplacian on the worldsheet,
\begin{equation}
\Delta=
-\frac{\partial^2}{\partial\tau^2}
-\frac{\partial^2}{\partial\sigma^2}.
\end{equation}
In this section, we assume that all the boundary conditions are Dirichlet
for simplicity.  Other boundary conditions can be treated similarly.

First, let us rederive the rectangle amplitude $I_{(D|D,D|D)}$ in
(\ref{zdddd})
with the path integral approach.
If the worldsheet is a rectangle of size
$a\times b$,
then $X$ has the mode expansion
\begin{equation}
 X(\sigma,\tau) = \sum_{m,n\geq 1} X_{mn}\sin\left(\frac{m\pi\tau}{a}\right)
\sin\left(\frac{n\pi\sigma}{b}\right).
\label{eigenmode}
\end{equation}
The eigenvalues of the operator $\Delta$ for
the eigenfunctions in (\ref{eigenmode}) are
\begin{equation}
\lambda_{m,n}=\frac{\pi^2m^2}{a^2}+\frac{\pi^2n^2}{b^2},\quad
m,n=1,2,\ldots.
\label{lambdamn}
\end{equation}
Then, the determinant of the operator $\Delta$ is represented
by the infinite product of all the eigenvalues
\begin{equation}
 I=(\det\Delta)^{-1/2}
 =\prod_{m,n=1}^\infty\lambda_{m,n}^{-1/2}.
\label{prodlambda}
\end{equation}
An advantage of the path integral approach is
that the modular invariance is manifest by definition.
Indeed, the amplitude (\ref{prodlambda}) is 
invariant under the exchange of $a$ and $b$.
However, this property has only a formal significance unless we regularize this
divergent product of the eigenvalues.

A simple way to obtain a finite result is
to use the $\zeta$-function regularization.
In this method, we first rewrite (\ref{prodlambda}) as
\begin{equation}\label{ptf1}
I=\prod_{m,n\geq1}\lambda_{m,n}^{-1/2}
=\prod_{m,n\geq 1}\frac{b}{\pi n}
\left(1+\frac{m^2t^2}{n^2}\right)^{-1/2},
\end{equation}
where $t=b/a$.
Let us carry out the product with respect to $n$.
Removing the divergent constant factor $\prod_{n=1}^\infty(1/\pi n)$ independent of
$a$ and $b$, and using the formula
$\prod_{n=1}^\infty(1+x^2/n^2)
=\sinh\pi x/(\pi x)$
for the convergent product and
the $\zeta$-function regularization formulae
$\prod_{n=1}^\infty x=x^{\zeta(0)}=x^{-1/2}$
and $\prod_{m=1}^\infty q^m=q^{\zeta(-1)}=q^{-1/12}$,
we obtain
\begin{equation}
I
\propto\prod_{m=1}^\infty b^{-1/2}\left(\frac{\sinh \pi mt}{\pi mt}\right)^{-1/2}
=\prod_{m=1}^\infty\left(\frac{aq^{-m}}{2\pi m}(1-q^{-2m})\right)^{-1/2}
\propto((ab)^{-1/4}t^{1/4}\eta(t))^{-1/2}\,,
\label{zbyzeta}
\end{equation}
where $q=e^{-\pi t}$.
The $t$-dependent part of this amplitude coincides with the
amplitude we obtained using the oscillator approach,
and the area dependence reproduces the
correct conformal weight, $-1/4$.

Although the $\zeta$-function regularization provides
a simple and efficient way to extract a finite result
from the divergent product,
its physical meaning is unclear.
For this reason, we give another derivation of the same result using a
more conventional regularization method, the heat kernel method,
in which we introduce a cut-off $\epsilon$ and regularize the determinant
by writing
\begin{equation}
\Gamma=\log\det\Delta
=-\int_\epsilon^\infty\frac{d\xi}{\xi}\tr e^{-\xi\Delta}\,,
\label{heatkernel}
\end{equation}
where we have defined $\Gamma$ through the relation $I=e^{-\Gamma/2}$.
Then, for the rectangle with eigenvalues
(\ref{lambdamn}), we obtain
\begin{eqnarray}
\Gamma
&=&
-\int_\epsilon^\infty\frac{d\xi}{\xi}\sum_{m,n}e^{-\xi\lambda_{m,n}}
=-\frac{1}{4}\int_\epsilon^\infty\frac{d\xi}{\xi}
\left(\theta_3\left(\frac{\pi\xi}{a^2}\right)-1\right)
\left(\theta_3\left(\frac{\pi\xi}{b^2}\right)-1\right)
\nonumber\\
&=&-\frac{ab}{4\pi\epsilon}
  +\frac{2(a+b)}{4\sqrt{\pi\epsilon}}
  -\frac{1}{4}\log\frac{ab}{\pi\epsilon}
  +\log(t^{1/4}\eta(t))
  +\frac{\gamma-\log 4\pi}{4}+{\cal O}(\epsilon).
\label{rectamp}
\end{eqnarray}
The first two terms here are cancelled by local counterterms
and have no physical meaning.
The remaining part correctly reproduces the previous result
(\ref{zbyzeta}),
up to a constant.

\subsection{Derivation of the rectangular diagram from the Weyl anomaly}
\label{sec:anomaly}
For a general worldsheet with boundaries,
a formula for the determinant defined with the regularization (\ref{heatkernel})
is given in Ref.~\citen{alvarez}.
This gives the determinant as an anomaly associated with the Weyl rescaling
$g_{ab}=e^{2\sigma}\wh g_{ab}$,
where $\wh g_{ab}$ is a fiducial reference metric.
Specifically, it gives the determinant as a functional of $\sigma$:
\begin{eqnarray}
\log I&=&-\frac{1}{2}(\Gamma_{\rm div}+\Gamma_{\rm bulk}+\Gamma_{\rm bdr}+\cdots)\,,
\nonumber\\
\Gamma_{\rm div}&=&-\frac{1}{4\pi\epsilon}\int d^2z\sqrt{\wh g}e^{2\sigma}
 +\frac{1}{4\sqrt{\pi\epsilon}}\int d\wh se^\sigma
+\frac{1}{6}\chi(M)\log\epsilon\,,
\nonumber\\
\Gamma_{\rm bulk}&=&
-\frac{1}{12\pi}\int_M\sqrt{\wh g}\wh g^{ab}(\partial_a\sigma)(\partial_b\sigma)\,,
\nonumber\\
\Gamma_{\rm bdr}&=&
 -\frac{1}{6\pi}\left(\int_{\partial M}d\wh s\wh k\sigma+\frac{1}{2}\int_Md^2z\sqrt{\wh g}\wh R\sigma\right)\,.
\label{weylanomaly}
\end{eqnarray}
Here, $\Gamma_{\rm div}$
represents the divergent part depending on the cutoff $\epsilon$,
and $\Gamma_{\rm bulk}$ and $\Gamma_{\rm bdr}$ are finite contributions
from the bulk and the boundary of the worldsheet, respectively.
(Although $\Gamma_{\rm bdr}$ includes the bulk integral term,
it vanishes in the computation carried out below.)
The dots represent terms independent of $\sigma$, in which we are not interested.
The first two terms in $\Gamma_{\rm div}$ are divergent terms proportional to
the area of the worldsheet and the length of the boundary, respectively.
They are invariant under the renormalization flow in the sense that
a rescaling of the cutoff, $\epsilon\rightarrow e^{2\alpha}\epsilon$,
is absorbed by the constant shift $\sigma\rightarrow\sigma+\alpha$.
The third term in $\Gamma_{\rm div}$, which is logarithmically divergent,
is also invariant under renormalization
if it is combined with $\Gamma_{\rm bdr}$.
The rescaling of the cutoff changes the third term of $\Gamma_{\rm div}$
by $(\alpha/3)\chi(M)$,
and this is cancelled by the change of $\Gamma_{\rm bdr}$
under the constant shift $\sigma\rightarrow\sigma+\alpha$,
\begin{equation}
\delta\Gamma_{\rm bdr}
=-\frac{\alpha}{6\pi}\left(\int_{\partial M}d\wh s\wh k+\frac{1}{2}\int_Md^2z\sqrt{\wh g}\wh R\right)=-\frac{\alpha}{3}\chi(M).
\label{rigid}
\end{equation}
In this way, logarithmically divergent terms are always accompanied by
finite terms which make the overall amplitude renormalization invariant.
Using this fact, we can determine the scale dependence of amplitudes
from their cutoff dependence.

For a rectangle of size $a\times b$, the first two terms
in $\Gamma_{\rm div}$
are identical to the first two terms in (\ref{rectamp}).
However, the logarithmically divergent term, which is
related to the conformal weight of the amplitudes,
does not coincide with the corresponding term in (\ref{rectamp}).
As we see from (\ref{rigid}),
the formula (\ref{weylanomaly}) gives a conformal weight $-1/6$ for any
amplitude in the case of disk topology, while we know that the rectangle amplitudes
have the conformal weight $-1/4$.
In order to resolve this problem, we have to treat the divergence of the
integral at the corners more carefully.

In order to see how the divergence at each corner modifies the
conformal weight, let us focus on one of corners.
We consider slightly generalized situation 
in which the corner has an angle of $\pi/n$.
We can always choose coordinates $z$ and $u$
for which
the mapping around the corner is represented by
$z\propto u^n$.
We employ the flat metric on the $z$-plane as the reference metric $\wh g$,
and $g$ is the flat metric on the $u$-plane.
From (\ref{rigid}),
we can extract the conformal weight
\begin{equation}
\lambda(\mbox{boundary contribution})=-\frac{n-1}{12n}
\label{curvcon}
\end{equation}
for the corner.
As mentioned above, this part is not sufficient
to reproduce the conformal weight of the corners.
For this purpose, we also need to take account of the contribution of $\Gamma_{\rm bulk}$.
For the mapping $z\propto u^n$, it is given by
\begin{equation}
\Gamma_{\rm bulk}
=-\frac{1}{12\pi}\left(\frac{n-1}{n}\right)^2
    \int d^2z\frac{1}{|z|^2}.
\label{eq70}
\end{equation}
This integral diverges logarithmically in the limit $|z|\rightarrow0$.
(Note that the divergence in the limit $z\rightarrow\infty$ is of no concern here.
Here we are focusing on the divergence at the corner.)
In order to regularize this divergence, we introduce a small cutoff $\rho$,
remove a sector of radius $\rho$ from the corner, and
integrate over only the region in the $z$-plane
mapped to values of $u$ satisfying $|u|\geq\rho$.
We then have
\begin{equation}
\Gamma_{\rm bulk}
=\frac{(n-1)^2}{12n}\log \rho+\mbox{($\rho$ independent part)}.
\label{gbulk}
\end{equation}
This cutoff dependence gives a new contribution to the
weight,
\begin{equation}
\lambda(\mbox{bulk contribution})=-\frac{(n-1)^2}{24n}\,.
\label{rec}
\end{equation}
Adding (\ref{curvcon}) and (\ref{rec}),
we obtain the correct value appearing in (\ref{lambdaeff}).

We can reproduce not only the conformal weight of a corner
but also the full rectangle amplitude as a Weyl anomaly.
For example, we can obtain a rectangle of arbitrary size
$a\times b$ from a round disk of diameter $d$.
Because the conformal map is singular at the corners of the
rectangle and the bulk integral term diverges there,
we remove small sectors around each corner.
Let $\rho_i$ ($i=1,2,3,4$) denote the radii of the sectors
defined with the metric on the rectangle.
Then we obtain the following conformal anomaly associated with the conformal map:
\begin{equation}
\Gamma=\log t^{1/4}\eta(t)-\frac{1}{4}\log(ab)
+\frac{1}{3}\log d-\frac{1}{6}\log 2+\frac{1}{2}\log\pi
+\frac{1}{24}\sum_{i=1}^4\log\rho_i\,.
\label{disk2rect}
\end{equation}
Up to constant terms, this coincides with the rectangle amplitude we have obtained with
other methods.
Note that the dependence of (\ref{disk2rect}) on each cutoff
is the same as that of (\ref{gbulk}).

The important point here is that
when the worldsheet with a corner is mapped to a worldsheet with a smooth boundary,
the image of a corner is `marked',
because we need to regularize the bulk term by, for example, removing a small
sector around each corner.
Mapping a rectangle to a round disk,
we obtain a disk with four marked points.

For this reason, the disk does not have conformal Killing vectors, and
it has a single modulus $t$.
When we compute the string amplitude,
we have to insert a ghost operator at the four marked points,
and we have to carry out a modulus integration
accompanied by a $b$-field insertion.

\section{Correlation functions}\label{example.sec}

In the previous sections, we have established the open boundary state by
demonstrating that it is identical to that derived with the path integral computation.
There, we learned that a special care is needed in the treatment of
the corner, especially when the vertex operator is inserted.  In this
section, we present  some examples of the
computation of the correlation functions in which
tachyon vertex operators are inserted at 
the corners. These examples illuminate the prescriptions we have proposed. 

\subsection*{\bf OBS with tachyon insertions}
We first consider the computation using OBS.
By analyzing the modular properties of the correlation function,
we reproduce the intrinsic weight of a corner of a rectangle
and the renormalized conformal weight
of the tachyon vertex operator at a corner 
discussed at the end of \S\ref{weight.sec}.
First, we consider a single free boson $X$ (i.e. the case $c=1$)
for simplicity.
Since the zero mode exists only in the mode expansion of NN string,
tachyon vertices can be inserted only at the corners
located between two edges with Neumann boundary conditions.
In order to evaluate amplitudes,
we must define the OBS with two tachyon vertices 
inserted and one with one tachyon vertex inserted.
The definition of the OBS with two tachyon vertices is expressed in terms of  
the renormalized tachyon vertex operators given in (\ref{rentach}):
\begin{eqnarray}\label{twotachren}
|B^o;k_1,k_2\rangle^N_{NN}
&\equiv&\lim_{\tau_1,\tau_2\rightarrow0}
V^{\rm ren}_{k_1}(\tau_1,0)
V^{\rm ren}_{k_2}(\tau_2,\pi)|B^o\rangle^N_{NN}
\nonumber\\
&=&2^{\frac{\wt k_1\wt k_2}{2}}
e^{-i(k_1+k_2)\wh x}\prod_{m=1}^\infty\exp\left(-\frac{\wt k_1+(-)^m\wt k_2}{m}a_{-m}\right)
|B^o\rangle^N_{NN}\,.
\end{eqnarray}
The definition of the OBS with one tachyon vertex is similar to 
that in the two vertex case considered above:
\begin{eqnarray}\label{onetachren}
|B^o;k\rangle^N_{ND}
&\equiv&\lim_{\tau\rightarrow 0}
V^{\rm ren}_{k}(\tau,0)|B^o\rangle^N_{ND}
=e^{-ik\wh x}\prod_{r=1/2}^\infty\exp
\left(-\frac{\wt k}{r}a_{-r}\right)|B^o\rangle^N_{ND}\,.
\end{eqnarray}
Here we have defined the dimensionless momentum 
$\wt k \equiv 2\sqrt{2\alpha^{\prime}}k$. Using $\wt k$,
the on-shell momentum for a tachyon is $\wt k^2=8$.

\begin{figure}[htb]
\begin{center}
\scalebox{1.0}[1.0]{\includegraphics{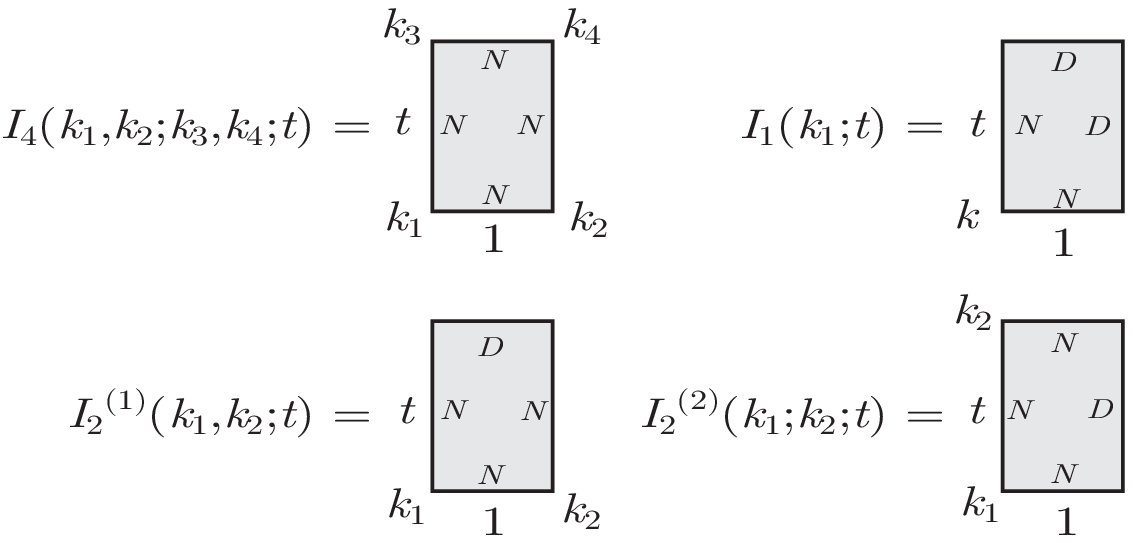}}
\end{center}
\caption{Rectangle amplitudes with tachyon vertex operators, specified by the momenta $k_i$.}
\label{amp.eps}
\end{figure}

\subsection*{\bf Four tachyon insertions}
When four edges of the rectangle have Neumann boundary conditions,
we can insert tachyon vertices at all the corners
(see Fig.\ref{amp.eps}):
\begin{eqnarray}
&&I_4(k_1,k_2;k_3,k_4;t)\label{four}
=
{}^N_{NN}\langle B^o;k_3,k_4|
q^{L^{NN}_0-\frac{1}{24}}
|B^o;k_1,k_2\rangle^N_{NN}\nonumber\\
&&~~~~~=
2^{\frac{\wt 1\wt 2+\wt 3\wt 4}{2}}\eta(t)^{\frac{1}{2}(\wt1^2+\wt2^2+\wt3^2+\wt4^2)-\frac{1}{2}}
\left(\frac{\eta(t/2)}{\eta(t)}\right)^{(\wt1-\wt2)(\wt3-\wt4)}
\left(\frac{\eta(2t)}{\eta(t)}\right)^{(\wt1-\wt3)(\wt2-\wt4)}.
\end{eqnarray}
(Henceforth we often use the notation $\wt i \equiv \wt k_i$.)
\begin{figure}[htb]
\begin{center}
\scalebox{1.0}[1.0]{\includegraphics{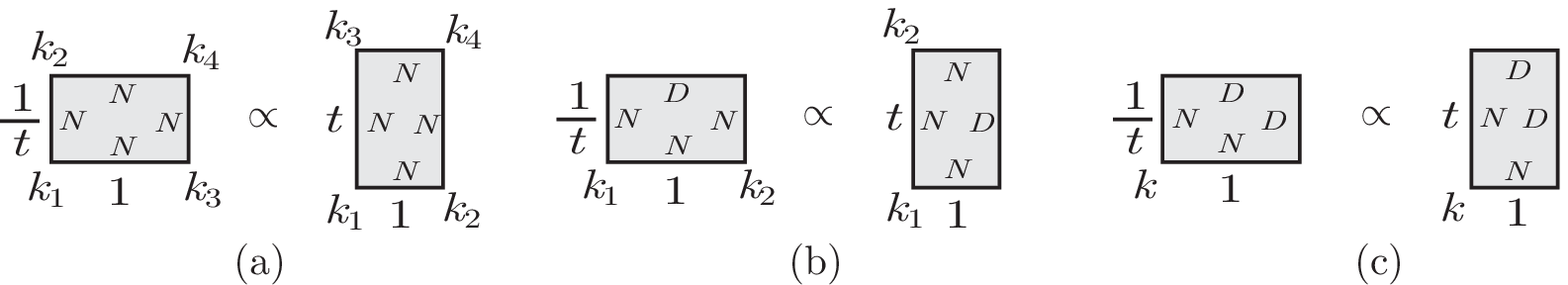}}
\end{center}
\caption{Modular properties of rectangle amplitudes with tachyon vertices}
\label{modular.eps}
\end{figure}
Using a computation similar to that yielding (\ref{rectymodular}),
the modular property of this amplitude is obtained as
\begin{equation}
I_4\left(k_1,k_2;k_3,k_4;\frac{1}{t}\right)
=t^{\frac{1}{4}(\wt 1^2+\wt 2^2+\wt 3^2+\wt 4^2-1)}
I_4(k_1,k_3;k_2,k_4;t)\,.
\end{equation}
This property is graphically depicted in Fig.\ref{modular.eps} (a).

The factor $t^{\frac{1}{4}(\wt 1^2+\wt 2^2+\wt 3^2+\wt 4^2-1)}$
on the right-hand side shows that the conformal weight of the entire
amplitude is 
$\frac{1}{4}(\wt 1^2+\wt 2^2+\wt 3^2+\wt 4^2-1)$.
This implies that the conformal weight of each 
tachyon vertex operator at a corner 
of the rectangle is $\wt k^2/4=2\alpha^{\prime} k^2$. 
This is identical to the weight of the 
{\it renormalized} vertex operator at the corner, 
discussed at the end of \S\ref{weight.sec}. 
This result comes from the use of the OBS with renormalized vertex operators.
The factor $-1/4$ comes from the intrinsic weight of the corner
(\ref{w122}), where all edges satisfy Neumann conditions.

We now comment on the relation between the on-shell 
condition of the tachyon vertex and the modular weight.
The condition of modular invariance must be imposed on the total amplitude 
$I_{\rm tot}=I_{\rm ghost} \prod I_{\rm matter}$.
Here the index of the product runs over all spacetime dimensions,
and $I_{\rm ghost}$ is the same as that used above.
Hence, the total conformal weight $w_{\rm tot}$ is 
\begin{equation}
w_{\rm tot}=\frac{1}{4}(\wt 1^2+\wt 2^2+\wt 3^2+\wt 4^2-26)+\frac{1}{2}\,.
\end{equation}
In order for the amplitude 
$A=\int_0^{\infty}  I_{\rm tot}(t)\,dt$
to be modular invariant, $w_{\rm tot}$ must be equal to 2.
If all the tachyons are on-shell, 
the conformal weight of the tachyon vertex is $\wt k^2/8=1$,
and therefore $w_{\rm tot}$ is in fact equal to 2.
This implies the equivalence of
the on-shell condition for the tachyon vertex
and the modular invariance of the rectangle amplitude.

\subsection*{\bf Comparison with UHP amplitude}
We now give a different type of calculation of the amplitude using
a conformal map and confirm the consistency with the
results obtained in the previous subsection.
This equivalence of these two approaches seems rather nontrivial.

The four tachyon amplitude $I_4^{\rm rec}(k_i;t)$ is obtained
by combining the ghost amplitude $I_{\rm ghost}$ in (\ref{ghamp})
and the amplitudes in (\ref{four}) for $26$ bosons as
\begin{equation}
I_4^{\rm rec}(k_i;t)
=
2^{2(s-4)}\eta(t)^4
\left(\frac{\eta(t/2)}{\eta(t)}\right)^{4(t-u)}
\left(\frac{\eta(2t)}{\eta(t)}\right)^{4(s-u)}\,,
\end{equation}
where we have assumed the on-shell condition $k_i^2=1/\alpha'$ and
defined the Mandelstam variables
$s=2\alpha'(k_1+k_2)^2$,
$t=2\alpha'(k_1+k_3)^2$,
and $u=2\alpha'(k_2+k_3)^2$.

Then, note that the same amplitude can be
obtained as the correlation function in the upper half plane
as
\begin{equation}\label{four-on}
I_4^{\rm UHP}
=\langle
ce^{-ik_3X}(0)
 e^{-ik_1X}(B)
ce^{-ik_2X}(1)
ce^{-ik_4X}(\infty)
\rangle
=(1-B)^{s/2-2}B^{t/2-2}\, .
\end{equation}

The next task is to determine the relation between the modulus $t$ 
of the rectangle worldsheet and the cross ratio $B$ of the 4-point function.
The modulus $t$ is defined by
\begin{equation}
it=\int_{B}^{0}\frac{dx}{y}\,\bigg/\!\int_{B}^{1}\frac{dx}{y},\quad
y^2=x(x-1)(x-B).
\label{tandB}
\end{equation}
We can solve (\ref{tandB}) to obtain $B$ in terms of $t$.
Then, we find that $B$ and its derivative are given by
\begin{equation}
B=\frac{\eta^{16}(t/2)\eta^8(2t)}{\eta^{24}(t)},
\quad
\frac{dB}{dt}=
16\pi\frac{\eta^{16}(t/2)\eta^{16}(2t)}{\eta^{28}(t)}\,.
\end{equation}
Using these relations, we obtain
\begin{equation}
I_4^{\rm UHP}dB=16\pi I_4^{\rm rec}dt\,.
\end{equation}

\subsection*{\bf Two tachyon insertions}
When there are 
three edges with Neumann boundary conditions and an edge with
Dirichlet boundary conditions,
we can insert two tachyon vertices. The amplitudes in this case are 
given by (see Fig.\ref{amp.eps})
\begin{eqnarray}
I^{(1)}_2(k_1,k_2;t)&=&
{}^D_{NN}\langle B^o|
q^{L_0^{NN}-\frac{1}{24}}
|B^o;k_1,k_2\rangle^N_{NN}\nonumber\\
&=&
2^{\frac{\wt 1\wt 2}{2}} e^{-i(k_1+k_2)x}
\left(\frac{\eta(t)}{\eta(2t)}\right)^{\frac{1}{2}}
\left(\frac{\eta(2t)^2}{\eta(t)}\right)^{\frac{\wt 1^2+\wt 2^2}{2}}
\left(\frac{\eta(4t)^2 \eta(t)}{\eta(2t)^3}\right)^{\wt 1 \wt 2}\,,
\end{eqnarray}
\begin{eqnarray}
I^{(2)}_2(k_1;k_2;t)&=&
{}^N_{DN}\langle B^o;k_2|
q^{L^{ND}_0+\frac{1}{48}}
|B^o;k_1\rangle^N_{ND}\nonumber\\
&=&
e^{-i(k_1+k_2)x}
\left(\frac{\eta(t)}{\eta(t/2)}\right)^{\frac{1}{2}}
\left(\frac{\eta(t/2)^2}{\eta(t)}\right)^{\frac{\wt 1^2+\wt 2^2}{2}}
\left(\frac{\eta(t/4)^2 \eta(t)}{\eta(t/2)^3}\right)^{\wt 1 \wt 2}\,.
\end{eqnarray}
These are related by the modular transformation
\begin{equation}\label{two-mod}
I_2^{(1)}\left(k_1,k_2;\frac{1}{t}\right)
=2^{\frac{1}{4}-\frac{\wt 1^2+\wt 2^2}{2}}
t^{\frac{1}{4}(\wt 1^2+\wt 2^2)} 
I_2^{(2)}(k_1;k_2;t)\,,
\end{equation}
which is graphically depicted in Fig.\ref{modular.eps} (b).
We can extract the weight of the renormalized vertex 
from the term $t^{\frac{1}{4}(\wt 1^2+\wt 2^2)}$. 
The two-tachyon amplitudes 
(\ref{two-mod}) possess the exact modular property 
up to the extra factor $2^{\frac{1}{4}-\frac{\wt 1^2+\wt 2^2}{2}}$,
but this extra factor 
can be absorbed in the redefinition of the boundary state.
The weight of the renormalized vertices is then identified with
$\frac{1}{4}(\wt 1^2+\wt 2^2)$.  In this case, 
the total intrinsic weight vanishes, 
which is clear from the assignment of the boundary conditions.

\subsection*{\bf One tachyon insertion}
In the case that two adjacent edges have Neumann boundary conditions
and the other two edges have Dirichlet boundary conditions,
we can insert only one tachyon vertex
(see Fig.\ref{amp.eps}):
\begin{eqnarray}
I_1(k;t)&=&
{}^D_{DN}\langle B^o|
q^{L^{ND}_0+\frac{1}{48}}
|B^o;k\rangle^N_{ND} \nonumber\\
&=&
e^{-ikx}\left(\frac{\eta(2t)\eta(t/2)}{\eta(t)^2}\right)^{\frac{1}{2}}
\left(\frac{\eta(t)^5}{\eta(2t)^2\eta(t/2)^2}\right)^{\frac{\wt k^2}{2}}.
\end{eqnarray}
Its modular property is
\begin{equation}
I_1\left(k;\frac{1}{t}\right)=t^{\frac{\wt k^2}{4}}I_1(k;t)\, ,
\end{equation}
which is graphically depicted in Fig.\ref{modular.eps} (c). 
This reproduces the weight of the renormalized vertex
$\frac{\wt k^2}{4}$.
The intrinsic weight vanishes again in this case,
due to  the assignment of boundary conditions.

\section{Conclusion}
In this paper, we introduced the boundary state in the open string
channel for a free boson and a fermionic ghost.
Their inner product gives a string amplitude
on a rectangle.  We carefully studied the CFT near the corner carefully
and elucidated certain nontrivial features, such as the
correction of the weight of primary fields.

There are many important issues which we did not address in this paper.
One of the most important problems is to generalize our treatment to other CFT,
for example, the minimal model \cite{Behrend:1999bn}, 
the orbifold \cite{Billo:2000yb}, 
the Wess-Zumino model \cite{Schomerus:2002dc}, 
and of course superstrings \cite{r:BCFT,Polchinski:1987tu}.
OBS for the Majorana fermion 
 is somehow nontrivial, because the naive analog of the fermionic ghost,
$e^{\frac{\pm i}{2} \sum_{r>0} \left(\psi_{-r}\right)^2}|0\rangle$, 
is meaningless.  In order to resolve this problem,
we have to be careful in treating the singularity at the corner.
The behavior of fermion fields must be considered in order to determine
how two D-branes can intersect.

Another important problem is to determine the proper normalization of
the open boundary state.  For the boundary states of a closed
string, it is fixed by the modular property (Cardy condition)
 \cite{Cardy:1989ir} 
and can be interpreted as the boundary entropy \cite{Affleck:1991tk}
 in statistical mechanics or the tension of the D-brane in string theory 
 \cite{r:Polchinski}.
This is closely related to the generalization to
generic CFT and is a very important problem
in the study of string dynamics.

We are also interested in the classical
field configuration near the D-brane.  For a closed string,
it is known \cite{9912161} that classical supergravity
solutions near the D-brane can be constructed from the massless
part of the boundary state.
It may be thought that, just as in the closed string case,
we can reproduce the soliton profile for source D-branes dissolved
in higher-dimensional D-branes by extracting the massless part
in the OBS. We note that
an analysis similar in spirit to that described here is carried out in 
Refs.~\citen{0211250} and \citen{0410062} for D$3$-D($-1$) systems, 
although the idea of OBS is not introduced there. 
The authors of these works computed a disk amplitude that is equivalent to 
$\langle P|q^{L_0}|B^o\rangle$ in our language,
where $\langle P|$ is a massless state of open strings on D3-branes
and $|B^o\rangle$ is the OBS for D($-1$)-branes.
They showed that this amplitude actually reproduces
the instanton profile on D3-branes. 
If we can translate their computation into OBS language, it would support
the conclusion that OBS can be interpreted as describing a soliton 
configuration in gauge theory. This would provide further
confirmation of the relevance of OBS.

We finally mention that OBS appear in the context
of string field theory \cite{Isono:2005ie} (see also
Refs.~\citen{Gaiotto:2002kf} and \citen{Ilderton:2004xq}) 
and are used in defining the modular
dual description of Witten's string field theory.
They define the on-shell external state in such a formulation
and directly define the D-brane as a solution to
string field theory.  Determining the role played by the boundary state
as a solution to the second quantized theory continues
to be a challenging problem.

\vskip 5mm
\noindent{\bf Acknowledgements}  ~We would like to thank
T.~Kawano, I.~Kishimoto and Y.~Tachikawa for their interesting comments.
Y.I. is supported in part by a Grant-in-Aid for the Encouragement
of Young Scientists (\#15740140).
Y.M. is supported in part by Grant-in-Aid (\#16540232) from
the Japan Ministry of Education, Culture, Sports, Science and Technology.
H.I. is supported in part
by a JSPS Research Fellowship for Young Scientists.


\end{document}